\documentclass[preprint,showkeys,secnumarabic,amsfonts,showpacs,amsmath,amssymb]{revtex4}

\usepackage{epsfig}

\usepackage{euscript,graphicx,amssymb,subfigure}

\usepackage{amsfonts}
\usepackage{amsmath}
\usepackage{amssymb}
\usepackage{graphicx}
\usepackage{euscript}
\usepackage{color}
\usepackage{subfigure}

\begin{document}

\title{Geometric and thermodynamic properties in Gauss-Bonnet gravity}

\author{Hossein Farajollahi $^{1,2}$}
\email{hosseinf@guilan.ac.ir}
\author{Amin Salehi $^{1}$}

\affiliation{$^1$Department of Physics, University of Guilan, Rasht, Iran}
\affiliation{$^2$ School of Physics, University of New South Wales, Sydney, NSW, 2052, Australia}

\begin{abstract}
In this paper, the generalized second law (GSL) of thermodynamics and entropy is revisited in the context of cosmological models in Gauss-Bonnet gravity with the boundary of the universe is assumed to be enclosed by the dynamical apparent horizon. The model is best fitted with the observational data for distance modulus. The best fitted geometric and thermodynamic parameters such as equation of state parameter, deceleration parameter and entropy are derived. To link between thermodynamic and geometric parameters, the "entropy rate of change multiplied by the temperature" as a model independent thermodynamic state parameter is also derived. The results show that the model is in good agreement with the observational analysis.
\end{abstract}

\maketitle

\section{Introduction}\label{s:intro}

Recent observations of high redshift type Ia
supernovae, the surveys of clusters of galaxies, Sloan digital sky survey (
SDSS) and Chandra X--ray observatory reveal the universe accelerating expansion
and that the density of matter is very much less than the critical density \cite{rie98}. In addition, the
observations of Cosmic Microwave Background (CMB)
anisotropies indicate that the universe is flat and the total energy
density is very close to the critical one \cite{ben03}.

The observations strongly indicates that the universe
presently is dominated by a smoothly distributed and slowly
varying dark energy (DE) component. A dynamical equation of state ( EoS) parameter that is connected directly to the evolution of the energy density in the universe and indirectly to the expansion of the Universe can be regarded as a suitable parameter to explain the acceleration and the origin of DE \cite{teg05}--\cite{set07}. In scalar-tensor theories \cite{sah00}--\cite{noj04}, interaction of the scalar field with matter ( for example in chameleon cosmology) can be used to interpret the late time acceleration and smoothly varying EoS parameter \cite{set10}--\cite{dim05}.

Motivated by the black hole physics, it was realized that there is a profound connection
between dynamic and thermodynamic of the universe (see for example \cite{jac95}--\cite{cai05}). In particular, the validity of the GSL {\cite{dav87}} which state that entropy of the fluid
inside the horizon plus the entropy associated with the apparent horizon do not decrease
with time, has been the subject of many studies.

In order to differentiate between cosmological models, a sensitive
and robust geometric diagnostic for dark energy models is proposed by \cite{sah00} that makes use of "statefinder" parameters. It probes the expansion dynamics of the universe through higher derivatives of the expansion factor as a natural companion to the deceleration parameter. In \cite{farajo} the authors discussed GSL, entropy and geometric parameters, including statefinders in chameleon cosmology with bouncing behavior. They also introduced the "entropy rate of change multiplied by temperature" parameter as a geo-thermodynamic parameter which can be used to differentiate among cosmological models. The parameter, which defined in terms of the second derivative of the scale factor of the universe and relates the geometric properties of the cosmological models with the thermodynamic one, together with deceleration parameter is adopted to explain the dynamic of the universe. Here, in Gauss-Bonnet gravity, by best fitting the mode parameters with the observational data for distance modulus using the chi-squared method, we implement the same analysis as \cite{farajo}. The advantage of best fitting the model parameters with the observational data is to find a more realistic and physically motivated understanding of the results.

The paper is organized as follows: Section two is devoted to the model independent, thermodynamic formulation of the cosmological models in relation to the dynamical parameters. In section three, we derive the field equations for Gauss-Bonnet cosmological model. In section four we best fit the model with observational data and obtain constraint on the model parameters. A geo-thermodynamic study is presented in section five with a summary given in section six.

\section{GSL and entropy}

 According to the recent observational data from type Ia Supernovae in an accelerating universe, the enveloping surface should be the apparent
horizon rather than the event one \cite{zho07}. So, we assume that the universe is enclosed by the dynamical apparent horizon with the radius given by $R_{h}=\frac{1}{\sqrt{H^{2}}}$ in a flat FRW universe \cite{cai05}.

By the horizon entropy and temperature ,
the dynamics of the entropy on the apparent horizon is \cite{dav87},
\begin{eqnarray}\label{entropyAH}
\dot{S_{h}}=2\pi R_{h} \dot{R_{h}}.
\end{eqnarray}

Also, from the Gibbs equation, the entropy of the universe inside the horizon can be related
to its effective energy density and pressure in the horizon with,
\begin{eqnarray}\label{Gibb's eq.1}
TdS_{in}=p_{eff}dV+d(E_{in}),
\end{eqnarray}
where $S_{in}$ is the internal entropy within the apparent horizon and $p_{eff}$ is the effective pressure in the model. If there is no energy exchange between outside and inside of the apparent horizon, thermal equilibrium realizes that $T = T_{h}$. Hence, the expression for internal energy can be written as $E_{in} = \rho_{eff}V$, with $V = \frac{4}{3}\pi R_{h}^{3}$. From equation (\ref{Gibb's eq.1}), by using Friedmann equation in the cosmological models and doing some algebraic manipulations we find that the rate of change of the internal entropy, horizon entropy and total entropy are respectively,
\begin{eqnarray}
\dot{S_{in}}&=&12\pi^2 R_h^2H(1+\omega_{eff})(1+3\omega_{eff}),\label{sdotin}\\
\dot{S_{h}}&=&24\pi^2 R_h^2H(1+\omega_{eff}),\label{sdoth}\\
\dot{S}_{total}&=&36\pi^2 R_h^2H(1+\omega_{eff})^2.\label{sdott}
\end{eqnarray}
For the rate of change of the internal entropy, equation (\ref{sdotin}), we find that for an expanding universe with acceleration, $H>0$, and $-1 <\omega_{eff}<-1/3$ in quintessence era, $\dot{S_{in}}< 0$. On the other hand, in phantom era, $\omega_{eff}<-1$, and decelerating universe, $\omega_{eff}>-1/3$, we obtain $\dot{S_{in}}>0$. From equation (\ref{sdoth}), it can be seen that again in an decelerating expanding universe and when $\omega_{eff}>0$, then $\dot{S_{h}}\geq 0$. Otherwise, $\dot{S_{h}}\leq 0$. Finally, in equation (\ref{sdott}), the sign of the total rate of change of the entropy, $\dot{S_{t}}$, for an expanding universe is independent of EoS parameter.

\section{The Model}

We start with the action
\begin{eqnarray}\label{action}
S=\int[\frac{1}{2}R-\frac{1}{2}g^{\mu\nu}\nabla_\mu\phi\nabla_\nu\phi-V(\phi)-f(\phi)G]\sqrt{-g}dx^{4},
\end{eqnarray}
where $G$ is the Gauss-Bonnet invariant coupled with the scalar field $\phi$ and $V(\phi)$ is the potential. In the FRW cosmology the field equations for the metric and also scalar field are
\begin{eqnarray}\label{fried1}
3H^{2}=\frac{1}{2}\dot{\phi}^{2}+V+24H^{3}\dot{f},
\end{eqnarray}
\begin{eqnarray}\label{fried2}
2\dot{H}+3H^2=-\frac{1}{2}\dot{\phi}^{2}+V+8H^{2}\ddot{f}+16H\dot{f}(\dot{H}+H^{2}),
\end{eqnarray}
\begin{eqnarray}\label{phiequation}
\ddot{\phi}+3H\dot{\phi}+V'+f'G=0,
\end{eqnarray}
where  dot means and prime mean derivative with respect to cosmic time and scalar field respectively and $G=24(\dot{H}H^2+H^4)$. From equations(\ref{fried1}) and (\ref{fried2}) one can define the effective EoS parameter as $\omega_{eff}\equiv \omega_{\phi}+\omega_{Gf}$, where
\begin{eqnarray}
\omega_{eff}=\frac{\frac{1}{2}\dot{\phi}^{2}-V-8H^{2}\ddot{f}-16H\dot{H}\dot{f}-16H^{3}\dot{f}}{\frac{1}{2}\dot{\phi}^{2}+V+24H^{3}\dot{f},}
\end{eqnarray}
and
\begin{eqnarray}
\omega_{fG}=\frac{-8H^{2}\ddot{f}-16H\dot{H}\dot{f}-16H^{3}\dot{f}}{24H^{3}\dot{f}}.
\end{eqnarray}
To study the thermodynamic behavior of the model at late time, the structure of the dynamical system is revisited by taking into account the following dimensionless variables,
\begin{eqnarray}
\chi=\frac{\dot{\phi}}{\sqrt{6}H},\zeta=\frac{V}{3H^2},\eta =H\dot{f}
\end{eqnarray}
and parameter
\begin{eqnarray}
\alpha=\frac{\ddot{f}}{\dot{f}H}
\end{eqnarray}
We then rewrite the field equations in terms of these variables as
\begin{eqnarray}
\frac{d\chi}{dN}&=&-\chi\frac{\dot{H}}{H^{2}}+\frac{\chi}{\sqrt{6}}\frac{\ddot{\phi}}{H^{2}}\label{fq1}\\
\frac{d\zeta}{dN}&=&-2\zeta\frac{\dot{H}}{H^{2}}+\sqrt{6}\beta\chi\zeta\\ \label{fq2}
\frac{d\eta}{dN}&=&\eta(\alpha+\frac{\dot{H}}{H^{2}})\label{fq3}
\end{eqnarray}
where
\begin{eqnarray}
\frac{\dot{H}}{H^{2}}&=&\frac{-3\chi^{2}-4\eta+4\eta\alpha}{1-8\eta}\\ \nonumber
\frac{\ddot{\phi}}{H^{2}}&=&-3\sqrt{6}\chi-3\beta\zeta-\frac{24\eta}{\sqrt{6}\chi}(\frac{\dot{H}}{H^{2}}+1).
\end{eqnarray}
Also, the Hamiltonian constraint, (\ref{fried1}), becomes
\begin{eqnarray}
\chi^{2}+\zeta+8\eta=1.\label{hamilcon}
\end{eqnarray}
In order to close the system of equations we make the following ansatz: We consider that $f(\phi)=f_{0}\exp{(\alpha\phi)}$    $V(\phi)=V_{0}\exp{(\beta\phi)}$ where $\alpha$ and $ \beta $ are  dimensionless constants characterizing the slope of potential $f(\phi)$ and $V(\phi)$. There are no priori physical motivation for these choices, so it is only purely phenomenological which leads to the desired behavior of phantom crossing. Using the constraint equation (\ref{hamilcon}) the three first order coupled nonlinear differential equations reduces to two equations for the new dynamical variables $\chi$ and $\zeta$. In the next section we solve the equations by best fitting the model parameters  $\alpha$, $\beta$ and initial conditions $\chi(0)$, $\zeta(0)$, and $H(0)$ with the observational data for distance modulus using the $\chi^2$ method. The advantage of simultaneously solving the system of equations and best fitting the model parameters is that the solutions become physically meaningful and observationally favored.

\section{Cosmological constraints}

The difference between the absolute and
apparent luminosity of a distance object is given by, $\mu(z) = 25 + 5\log_{10}d_L(z)$ where the luminosity distance quantity, $d_L(z)$ is given by
\begin{equation}\label{dl}
d_{L}(z)=(1+z)\int_0^z{-\frac{dz'}{H(z')}}.
 \end{equation}
With numerical computation, we solve the system of dynamical equations for $\chi$, $\zeta$ and $\Omega_{V}$. While best fitting the model parameters and initial conditions with the most recent observational data, the Type Ia supernovea (SNe Ia), in order to accomplish the mission, we need the following two auxiliary equations for the luminosity distance and the hubble parameter
\begin{eqnarray}
 \frac{dH}{dN}=H(-\frac{\dot{H}}{H^{2}}),
 \end{eqnarray}
\begin{eqnarray}
\frac{d(d_{L})}{dN}=-(d_{L}+\frac{e^{-2N}}{H}).
\end{eqnarray}
To best fit the model for the parameters $\alpha$, $\beta$ and the initial conditions $\chi(0)$, $\zeta(0)$ and $H(0)$ with the observational data, SNe Ia, we employe the $\chi^2$ method. We constrain the parameters including the initial conditions by minimizing the $\chi^2$ function given as
\begin{eqnarray}\label{chi2}
&\chi^2_{SNe}&(\alpha,\beta;\chi,\zeta,H|_0)\nonumber\\
&=&\sum_{i=1}^{557}\frac{[\mu_i^{the}(z_i|\alpha,\beta;\chi,\zeta,H|_0)-\mu_i^{obs}]^2}{\sigma_i^2},
\end{eqnarray}
where the sum is over the SNe Ia data. In relation (\ref{chi2}), $\mu_i^{the}$ and $\mu_i^{obs}$ are the distance modulus parameters obtained from our model and observation, respectively, and $\sigma$ is the estimated error of the $\mu_i^{obs}$. Table I shows the best fitted model parameters and initial conditions.\\

\begin{table}[ht]
\caption{Best-fitted model parameters} 
\centering 
\begin{tabular}{c c c c c c c} 
\hline 
  &  $\alpha$  &  $\beta$ \ & $\chi(0)$\ & $\zeta(0)$
\ & $\chi^2_{min}$\\ [2ex] 
\hline 
&$3.96$  & $0.62$ \ & $0.6$\ & $-3.36$\ &   $548.0229449$ \\
\hline 
\end{tabular}
\label{table:1} 
\end{table}\

Figs. 1 shows the best-fitted one dimensional likelihood for the model parameters $\alpha$, $\beta$. In Fig. 2, the two dimensional likelihood and the confidence level at the $68.3\%$, $95.4\%$ and $99.7\%$ are shown.

\begin{figure}[t]
\includegraphics[scale=.3]{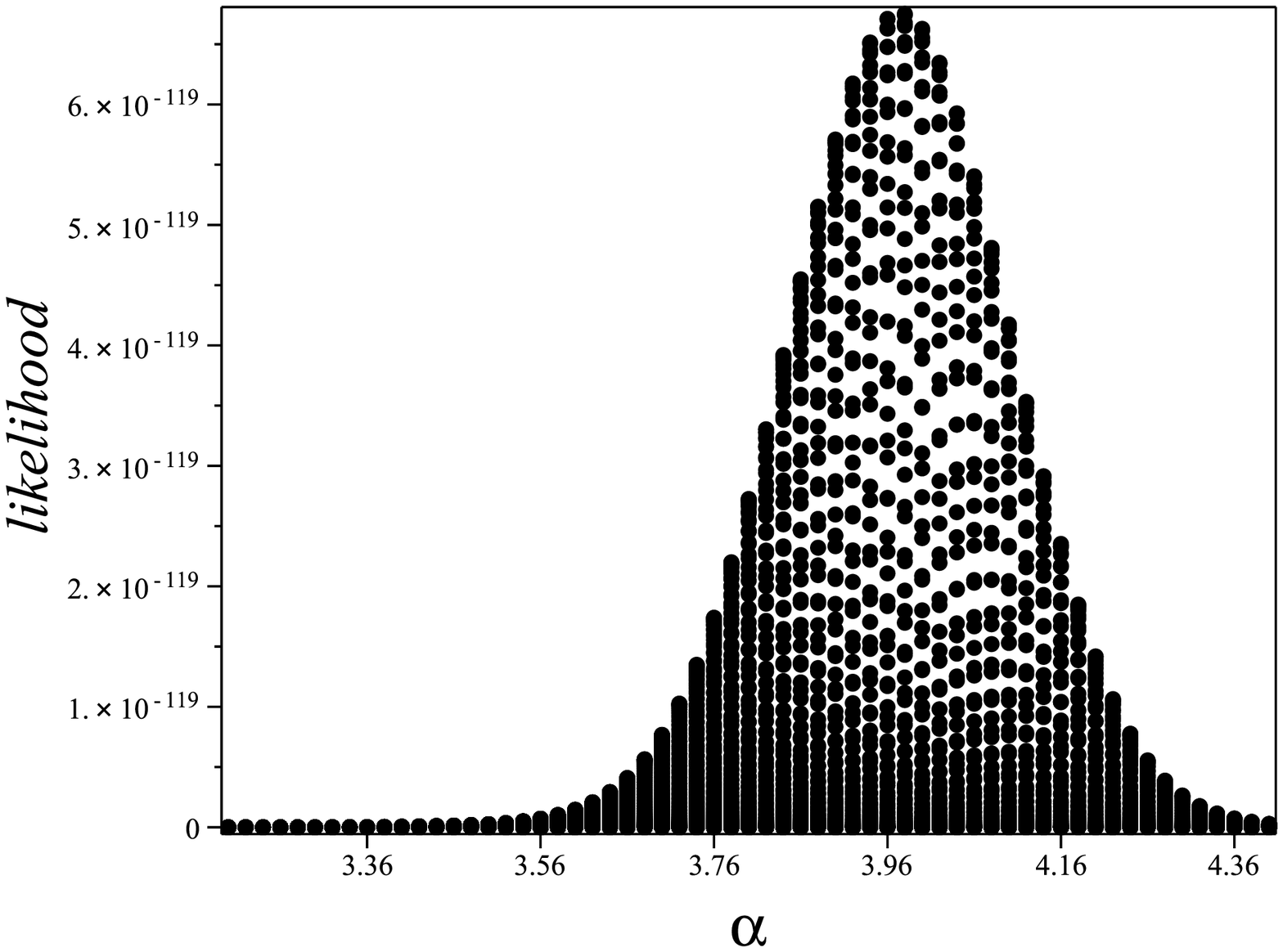}\hspace{0.1 cm}\includegraphics[scale=.3]{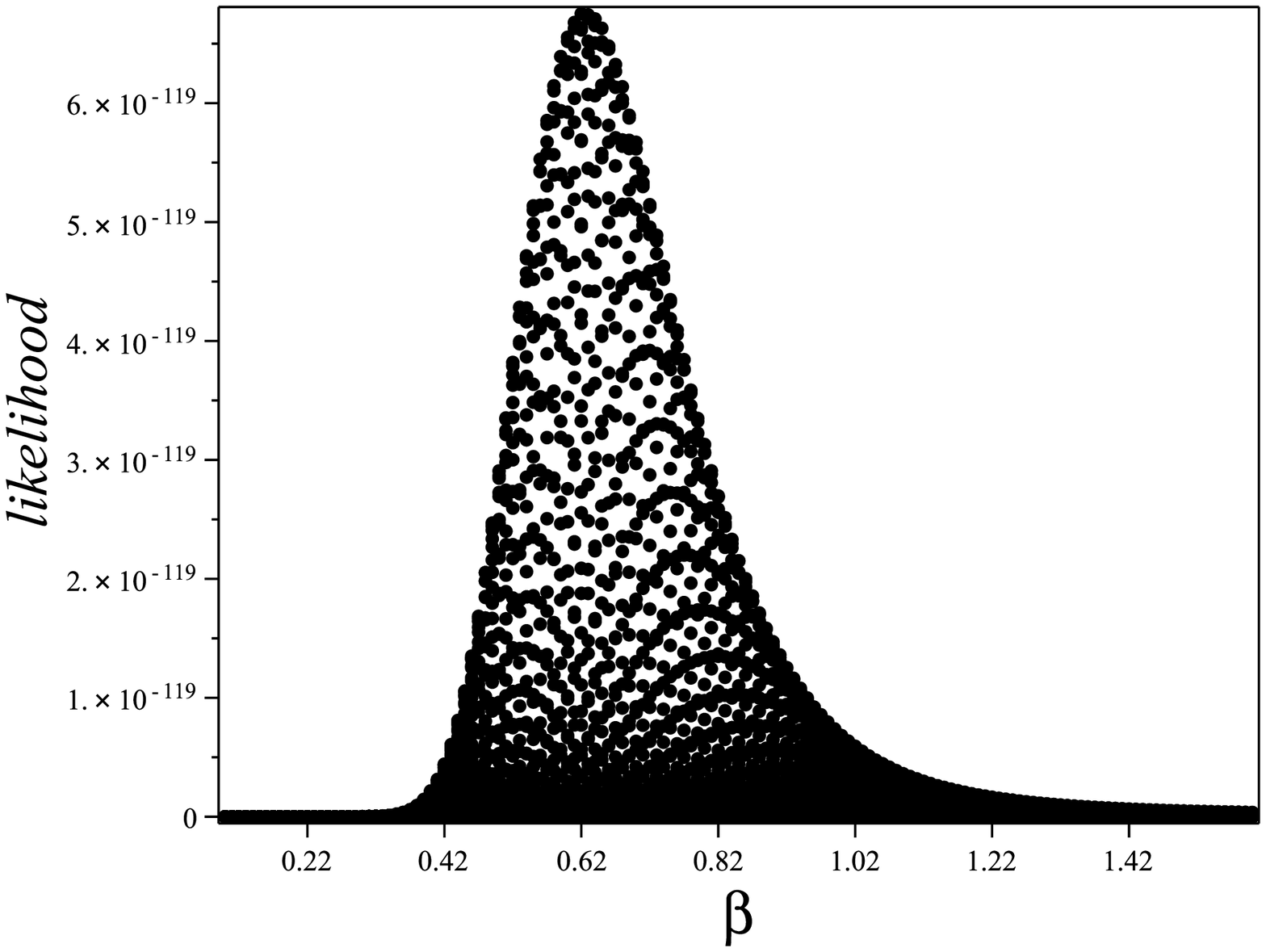}\hspace{0.1 cm} \\
Fig. 1:  The best-fitted one dimension likelihood for parameters $\alpha$ and $\beta$.
\end{figure}

\begin{figure}[t]
\includegraphics[scale=.3]{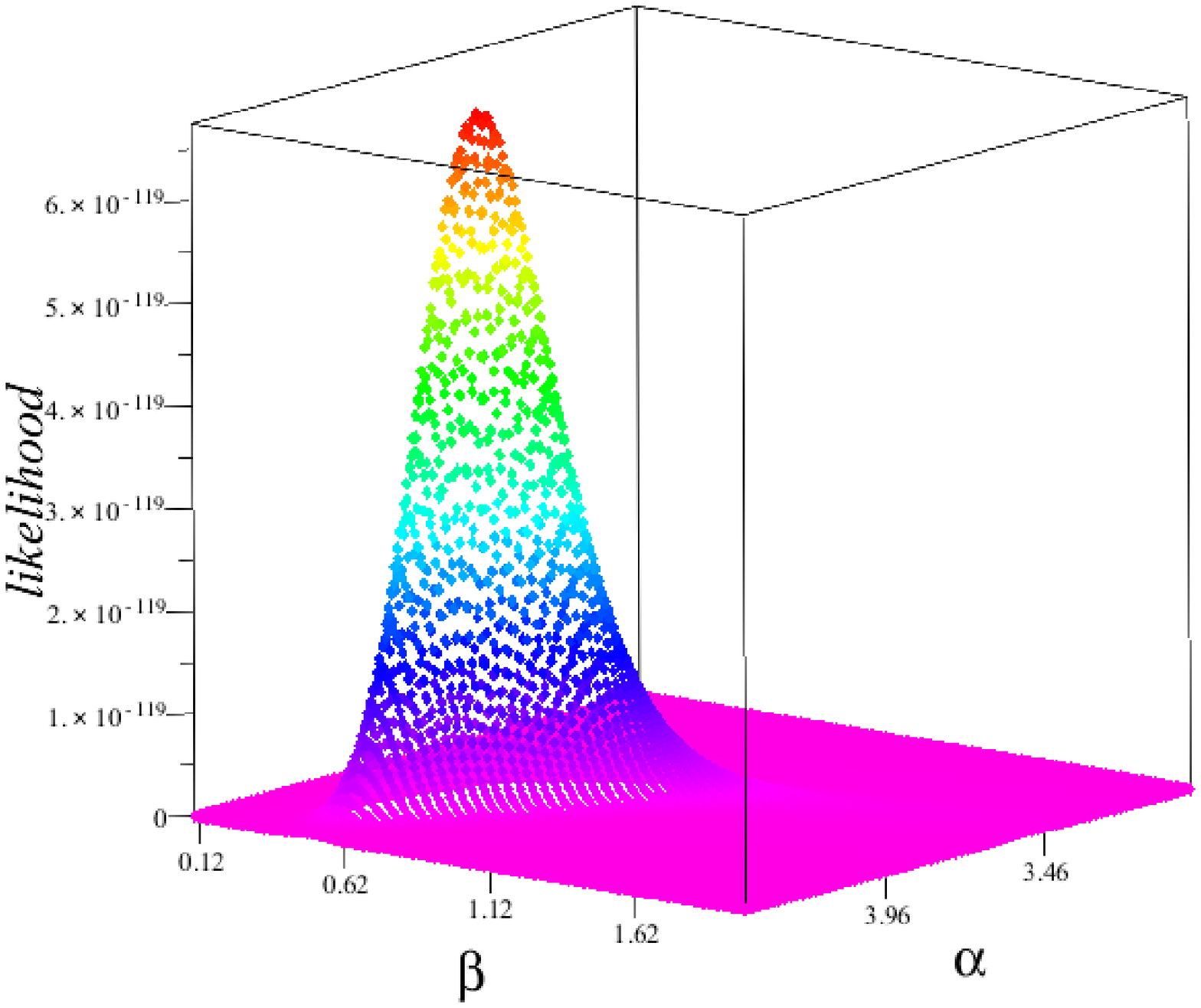}\hspace{0.1 cm}\includegraphics[scale=.3]{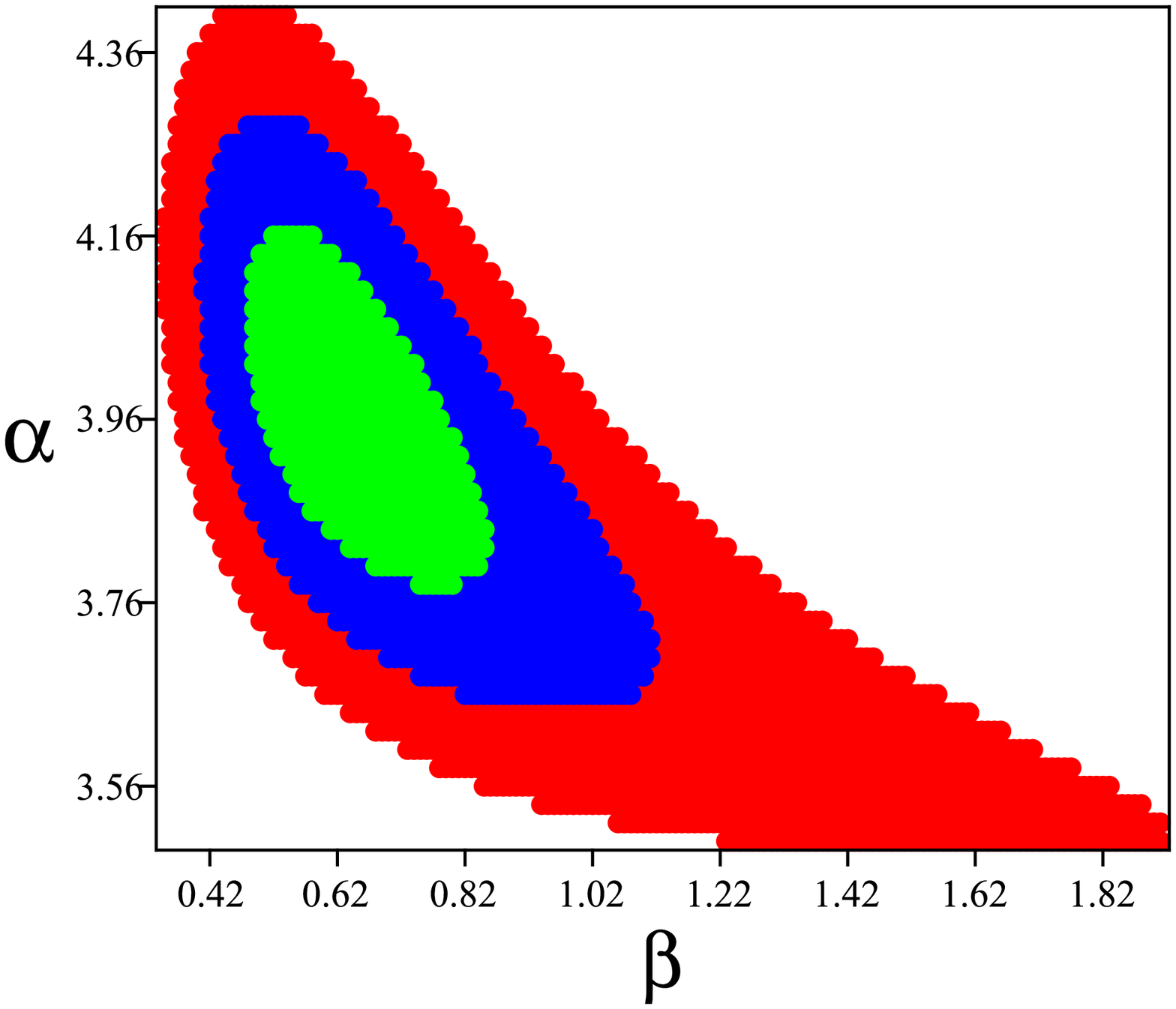}\hspace{0.1 cm}\\
Fig. 2:  The best-fitted two dimension likelihood and confidence level for parameters $\alpha$ and $\beta$.
\end{figure}

In Fig. 3, the distance modulus, $\mu(z)$, in our model is fitted with the observational data for the model parameters $\alpha$, $\beta$ and initial conditions for $\chi(0)$, $\zeta(0)$ and $H(0)$ using $\chi^2$ method in both cases of power law and exponential functions.

\begin{figure}[t]
\includegraphics[scale=.3]{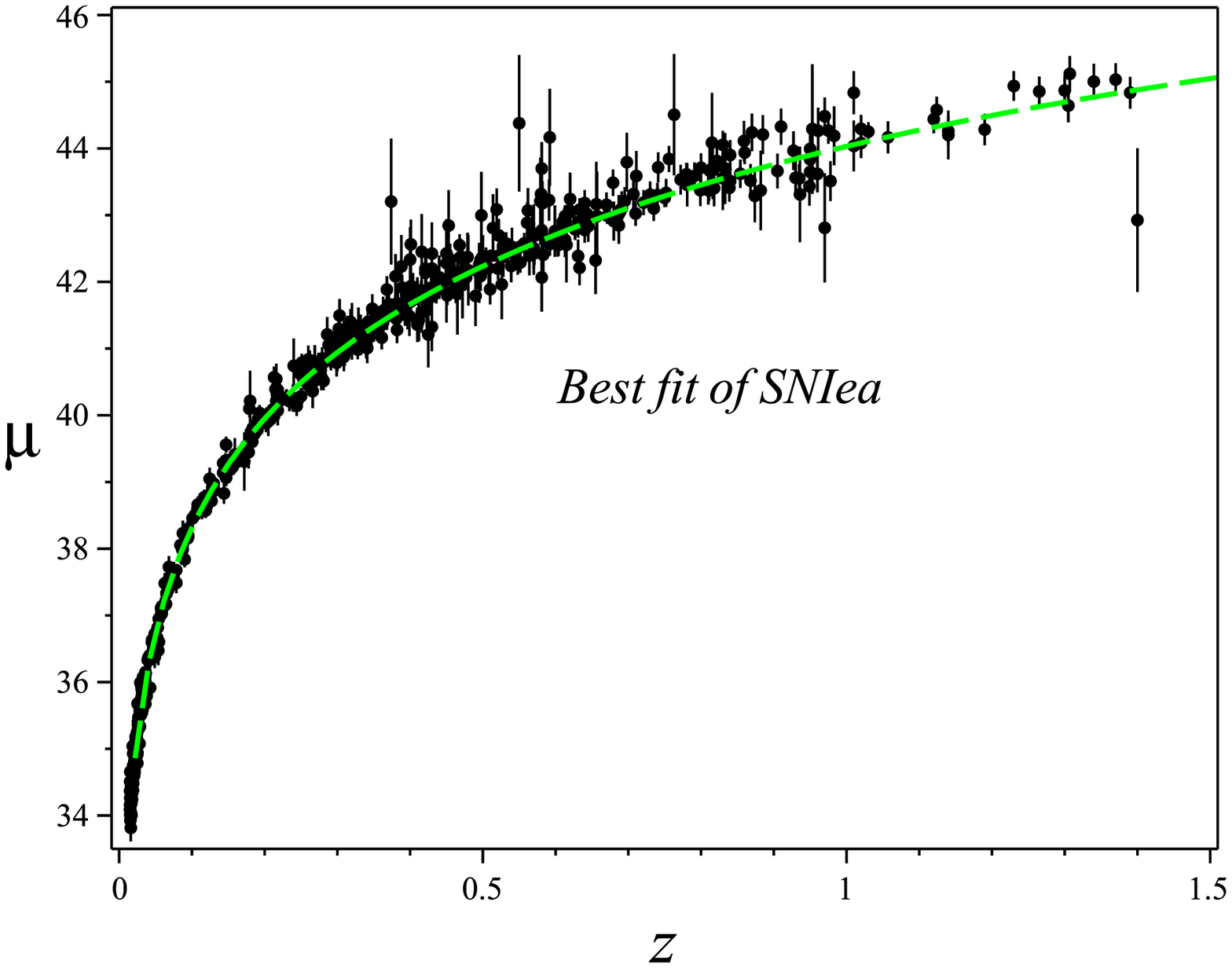}\hspace{0.1 cm}\\
Fig. 3:  The best-fitted distance modulus $\mu(z)$ plotted as function of redshift.
\end{figure}

Next, we examine our numerically solved and best fitted model with some other observational analysis. The geometric parameters such as EoS parameter and deceleration parameter together with thermodynamic parameters like entropy will be discussed. In addition, a description of the model is given by introducing the "total entropy rate of change, $\dot{S}_{total}$, multiplied by temperature $T$" as a physical variable.

\section{Geo-thermodynamic study}

In order to understand the behavior of the universe and its dynamics we need to study the cosmological parameters such as EoS and deceleration parameters. We have already verified our model with the current observational data via the distance modulus test. The EoS parameters analytically and/or numerically have been investigated by many authors for variety of cosmological models. The effective EoS parameter in terms of the new dynamical variables is given by
\begin{eqnarray}\label{eos}
\omega_{eff}&=&\frac{2\chi^{2}+\frac{8\eta}{3}(4-\alpha)-1}{1-8\eta}
\end{eqnarray}
With the best-fitted model parameters and initial conditions with the observational data, the effective EoS parameter is shown in Fig. 4) top. The graph shows that the universe starts from matter dominated era at higher redshift; enters the quintessence dominated era at $z =0.5$ where the universe begins accelerating in the near past. The result can be verified by Fig. 4)below for the best fitted deceleration parameter in which $q<0$ for $z<0.5$.

\begin{figure}[t]
\includegraphics[scale=.3]{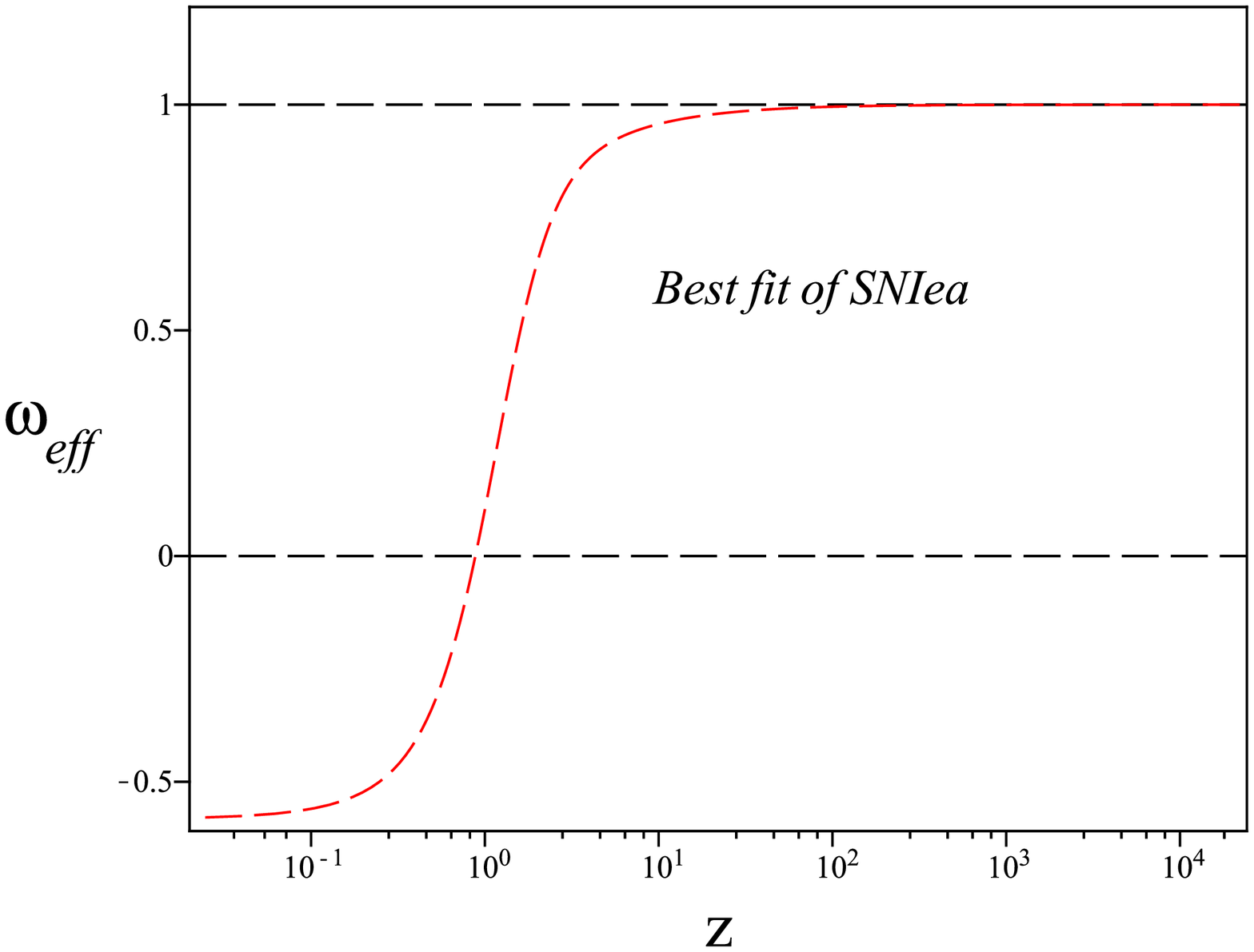}\hspace{0.1 cm}\includegraphics[scale=.3]{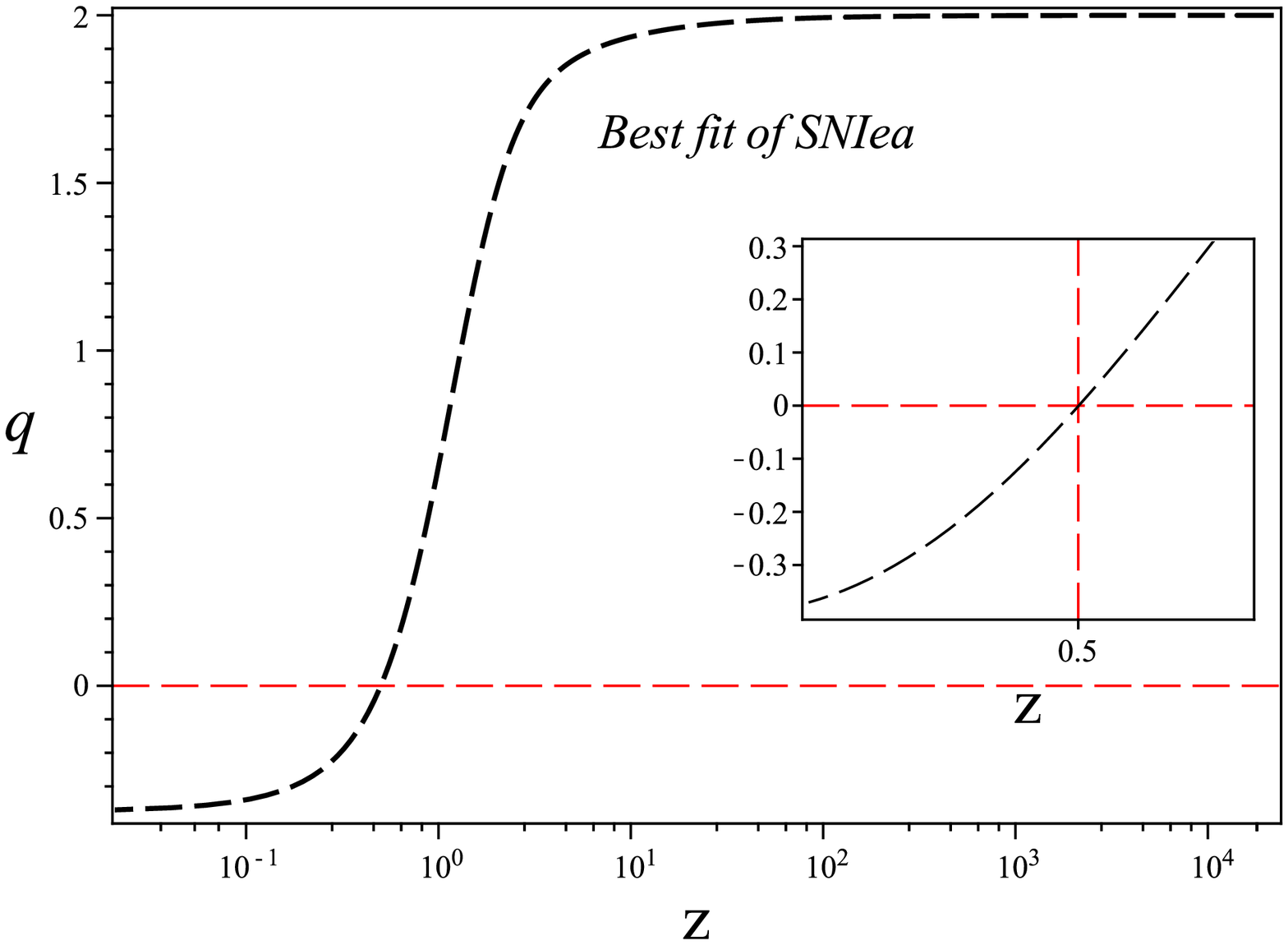}\hspace{0.1 cm}\\
Fig. 4:  The best-fitted effective $EoS$ and deceleration parameters plotted as function of redshift.
\end{figure}

In Fig.5, the dynamics of internal and total entropy rate of change are plotted. The graph for $\dot{S}_{in}$ shows that at $z>0.5$ where the universe begins accelerating and the effective EoS parameter becomes less than $1/3$, the internal entropy rate of change becomes negative as expected from eq. (\ref{sdotin}). Moreover, from the Fig. 5)below, we observe that the total entropy rate of change, $\dot{S}_{total}$ is always positive that can be verified by eq. (\ref{sdott}).

\begin{figure}[t]
\includegraphics[scale=.3]{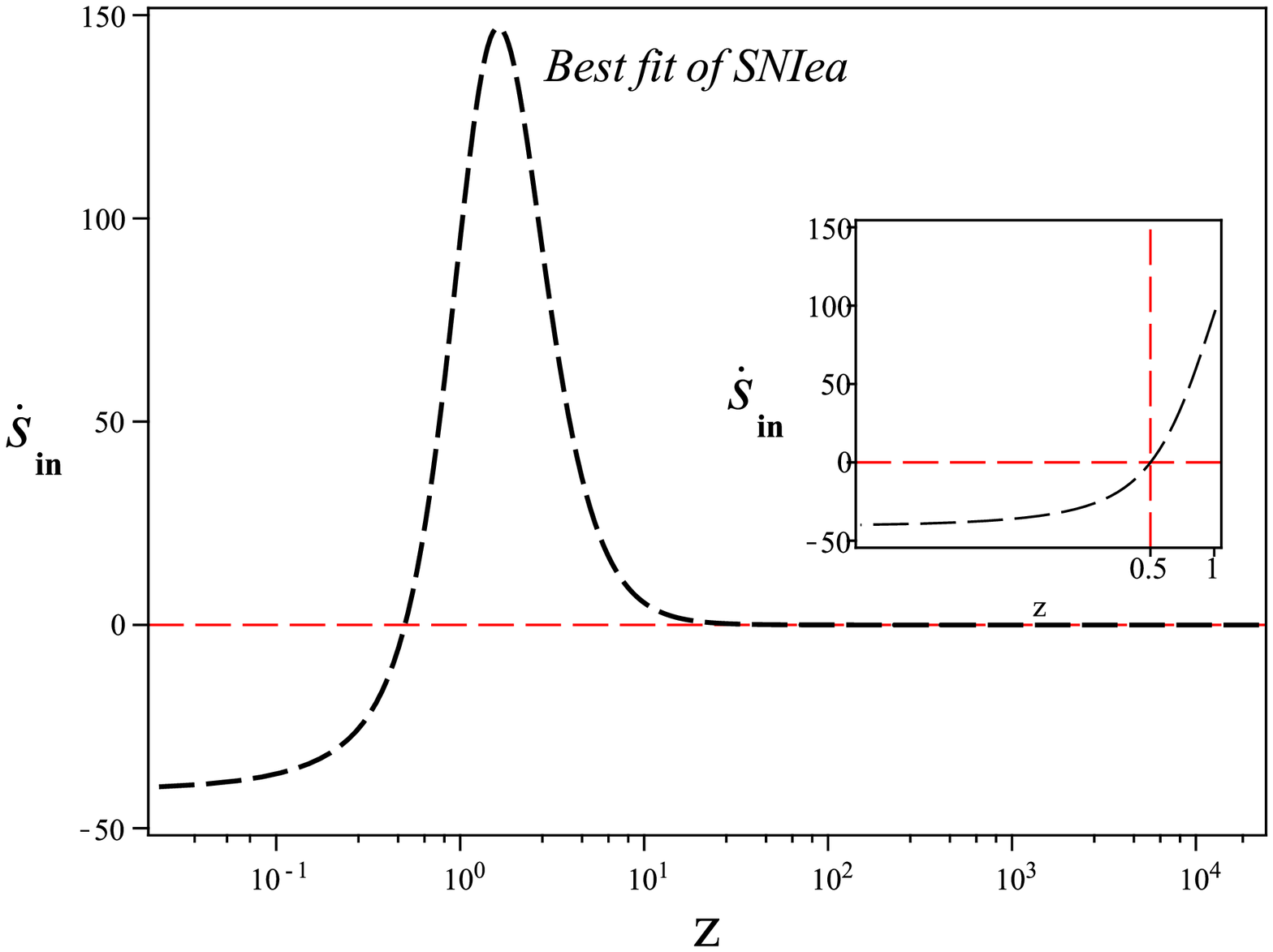}\hspace{0.1 cm}\includegraphics[scale=.3]{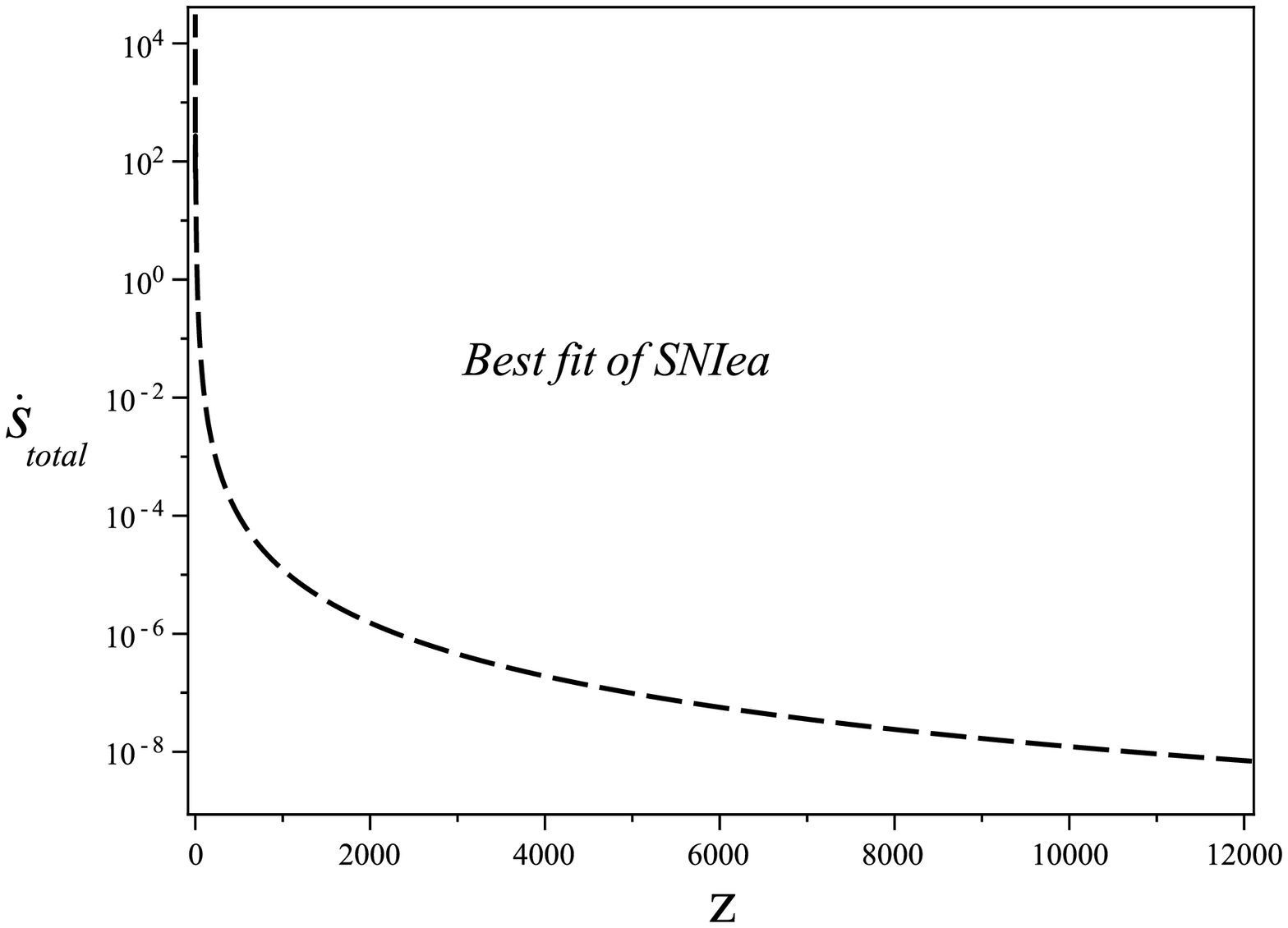}\hspace{0.1 cm}\\
Fig. 5:  The best-fitted internal and total entropy rate of change $\dot{S}_{in}$ and $\dot{S}_{total}$as a function of redshift.
\end{figure}

From the graphs, at higher redshifts, the EoS parameter is positive (matter dominated era) and the universe decelerates. The $\dot{S}_{total}$ gradually increases to its maximum value at $z=0$ whereas EoS parameter decreases to its minimum value. Motivated from \cite{farajo}, the "total entropy rate of change multiplied by temperature $T$" in terms of the new dimensionless dynamical variables and best fitted model parameters is given by
\begin{eqnarray}
u&=&\dot{S}_{total}T=8\pi(\frac{\dot{H}}{H^2})^2\nonumber\\
&=&8\pi(\frac{\dot{H}}{H^2})^2=8\pi[\frac{-3\chi^{2}-4\eta+4\eta\alpha}{1-8\eta}]^2.
\end{eqnarray}
The parameter $u$ is presented in terms of thermodynamic variables entropy and temperature or with the expression $\frac{\dot{H}}{H^2}$ which is a geometric quantity. We consider it as a geo-thermodynamic variable that relates the geometric properties of a cosmological model to its thermodynamic one. In Fig. 6, the best fitted $u$ is shown with respect to the redshift $z$. The dynamical behavior of $u$ is very similar to the dynamics of EoS and deceleration parameters. From the graph, in matter dominated era at higher redshift, $u$ is constant. At about $z\simeq 5$, it starts to decrease sharply until about $z\simeq 0.1$ in the near past when it become flat again.

\begin{figure}[t]
\includegraphics[scale=.3]{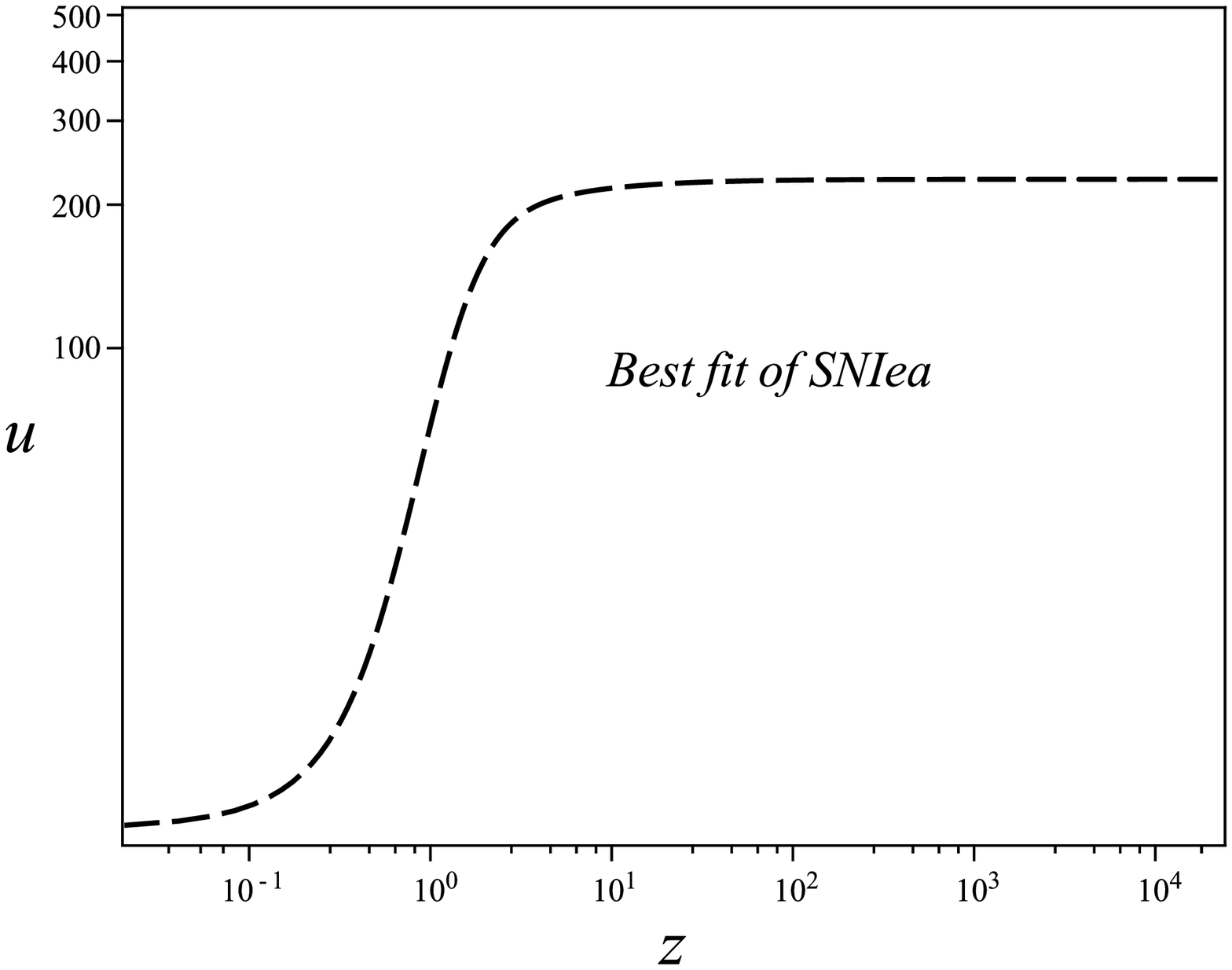}\hspace{0.1 cm}\\
Fig. 6: The best-fitted geo-thermodynamic variable $u$ as a function of redshift.
\end{figure}

In Fig. 7, we finally reconstruct the potential function $V(\phi)$, the coupled function $f(\phi)$ and also $\dot{\phi}$ against redshift $z$ for with the best fitted model parameters. As can be see, while the coupling function $f(\phi)$ is almost flat and low for $z<0$, it grows rapidly in near future. On the other hand, the negative potential $V(\phi)$ starts growing fast at very far redshift, $15000<z<22000$, but approaches zero later in the past till now. The velocity $\dot{\phi}$ also begins with large values in the far past and tends to zero in the near past till now.

\begin{figure}[t]
\includegraphics[scale=.3]{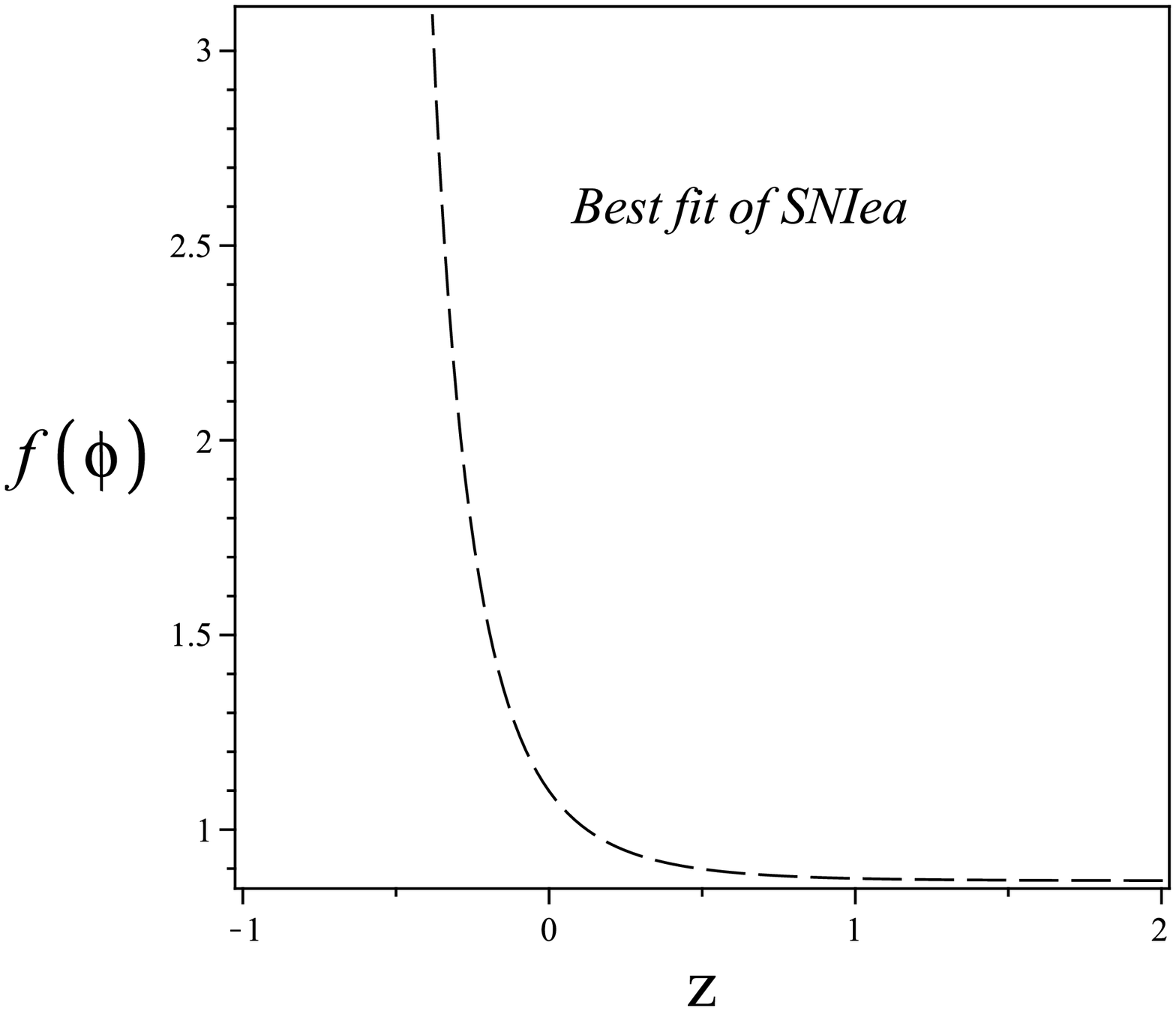}\hspace{0.1 cm}\includegraphics[scale=.3]{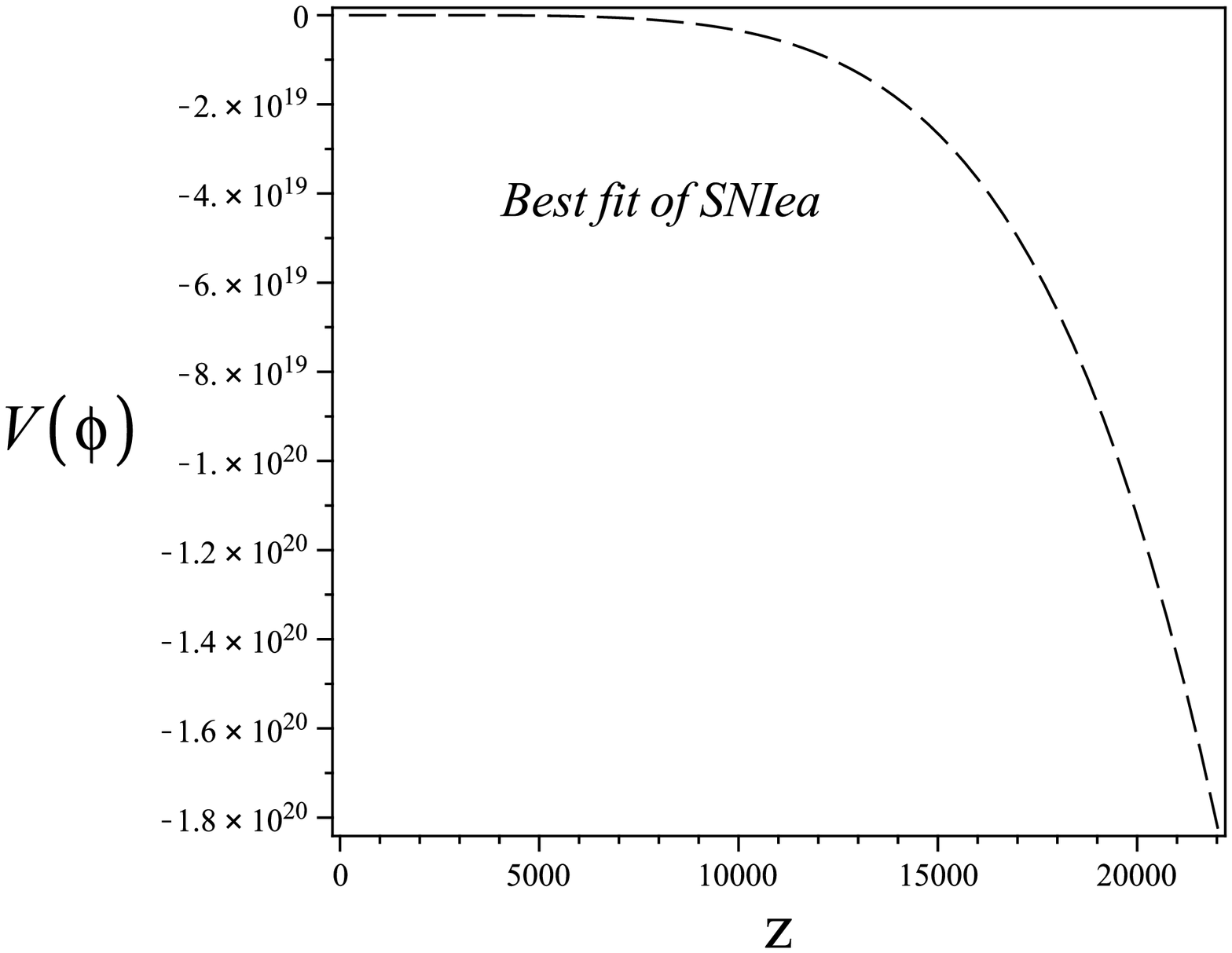}\hspace{0.1 cm}\includegraphics[scale=.3]{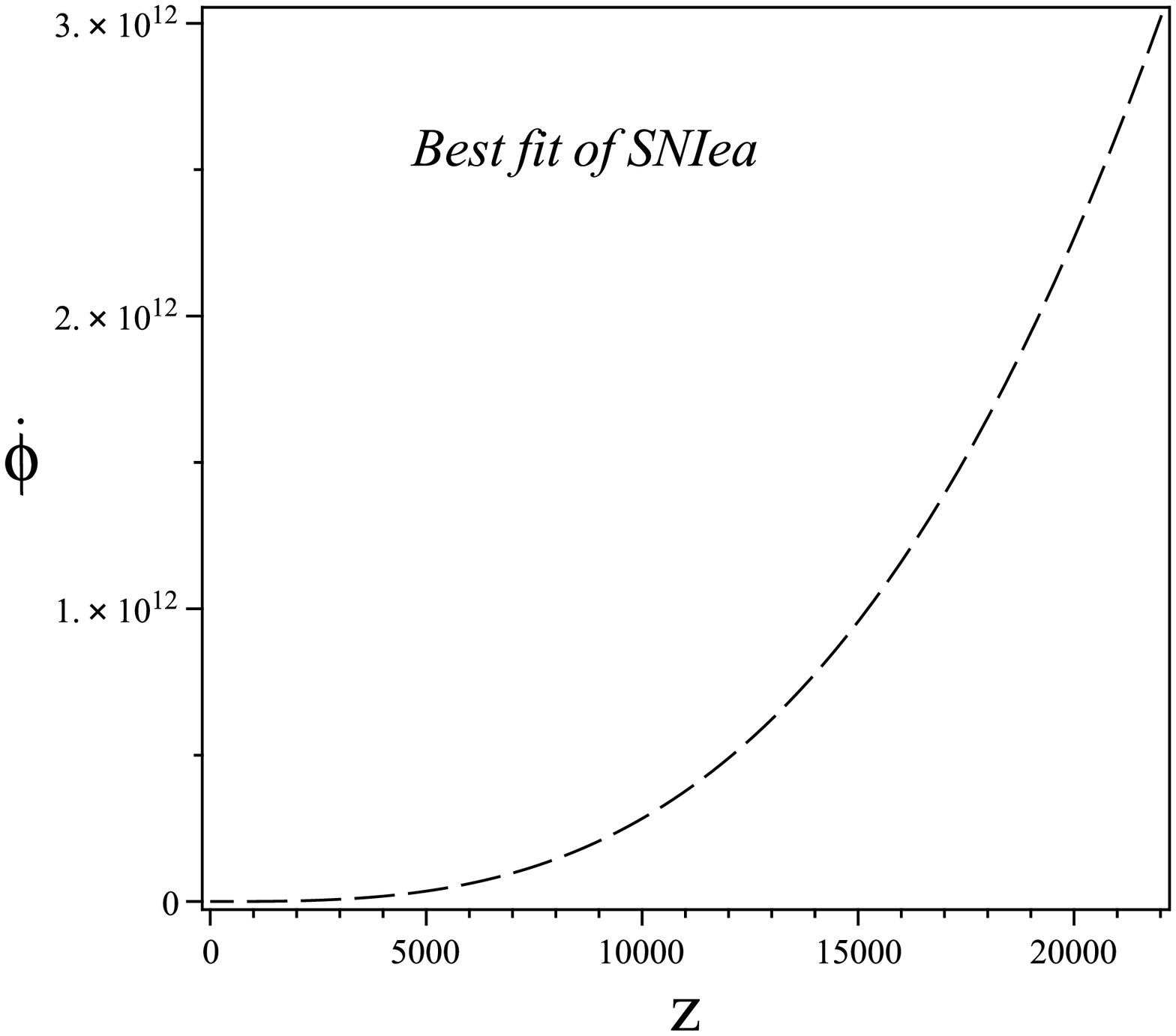}\hspace{0.1 cm}\\
Fig. 7:  The 1-dim and 2-dim likelihood distribution for parameters $\omega_{0}$ and $\beta$ in $\gamma=1/3$ case.
\end{figure}

\section{Summary}

In this paper we investigate the geometric and thermodynamic properties of the Gauss-Bonnet gravity. By best fitting the model parameters with the recent observational data for distance modulus using the chi-squared statistical method, we achieve a more reliable and physically motivated understanding of the results. From numerical calculation, the geometric dynamical variables such as EoS and deceleration parameters show that the universe transits from a period of matter dominated era and approaches a period of quintessence dominated era in the near past where the universe accelerate, but it never enters the phantom dominated era in the near past. A thermodynamic analysis shows that the rate of change of the total entropy of the universe is always positive which is consistent with the geometric findings. Furthermore, the rate of change of the internal entropy, when the universe enters quintessence dominated era (accelerating period), become negative as expected.

Following our previous work in \cite{farajo}, in a geo-thermodynamic prescription of the model, a new dynamical variable is derived for the model that inter-relate the geometric and thermodynamic properties of the cosmological models. Finally, we reconstructed the potential function, the coupled function and the velocity of the scalar field velocity in terms of best fitted model parameters. The results show that $f(\phi)$ has relatively more contribution to the late time acceleration than $V(\phi)$ and $\dot{\phi}$.


\begin{thebibliography}{}


\bibitem[(Riess 1998, Perlmutter 1999, Tonry 2003, Bennet 2003, Netterfield 2002, Halverson 2002, Pope 2004, Riess 2004,Knop 2003, Abazajian 2004, Tegmark 2004, Allen 2004)]{rie98} Riess, A. G. et al., 1998,Astrophys. J., 116, 1009; Perlmutter, S. et al., 1999, Astrophys. J. 517 565; Tonry, J. L.  et al., 2003, Astrophys. J.. 594, 1-24; Bennet, C. I. et al., 2003, Astrophys. J.s 148:1 ;
 Netterfield, C. B. et al., 2002, Astrophys. J. 571, 604;  Halverson, N. W. et al., 2002, Astrophys. J. 568, 38; Pope, A. C., et al., 2004, Astrophys. J. 607 655;  Riess, A.G. et al., 2004, Astrophys. J.  607 665; Knop, R. A. et al., 2003, Astrophys. J. 598, 102;  Abazajian, K. et al., 2004, Astron. J. 129,1755;2004 Astron. J. 128, 502; 2003 Astron. J. 126, 2081;
 Tegmark M. et al., 2004, Astrophys. J. 606, 702; Allen, S.W. et al., 2004, Mon. Not. R. Astron. Soc. 353, 457 (2004).

\bibitem[(e.g., Bennett et al. 2003, Spergel et al. 2003)]{ben03} Bennett, C. L. et al., 2003, Astrophys. J.s 148, 1;  Spergel, D. N. et. al., 2003, Astrophys. J.s 148 175.

\bibitem[(e.g., Tegmark 2005, Farajollahi et al. 2010,Farajollahi  \& Milani 2010, Farajollahi \& Mohamadi 2010, Seljak 2005)]{teg05}Tegmark, M.  2005, J. Cosmol. Astropart. Phys., 0504,001;  Farajollahi, H. et al., 2010, Mod. Phys. Lett. A, 25, 30 2579-2589;  Farajollahi, H. \& Milani, F. 2010, Mod. Phys. Lett. A25:2349-2362;  Farajollahi, H. \& Mohamadi, N. 2010, Int. J. Theor. Phys. 49:72-78; Seljak, U. et al, 2005, Phys. Rev. D 71, 103515.

\bibitem[(e.g. , Setare 2007, Sadeghi 2009, Farajollahi et al. 2011)]{set07}Setare, M. R. 2007, Phys. Lett. B644:99-103;  Sadeghi, J. et al., 2009,  Eur. Phys. J. C64:433-438; Sadeghi, J. et al., 2009, Phys. Lett. B679:302-305; Sadeghi,  J. et al., 2009,  Phys. Lett. B678:164-167; Farajollahi et al. 2011, Astrophys. Space Sci. DOI 10.1007/s10509-011-0779-6.



\bibitem[(e.g. , Sahoo \& Singh 2002, Sadeghi et al. 2009 )]{sah02} Sahoo, B.K. \&  Singh, L.P. 2002, Mod. Phys. Lett. A 17, 2409; Sahoo, B.K. \& Singh, L.P. 2003, Mod. Phys. Lett. A 18, 2725;  Sahoo, B.K. \& Singh, L.P. 2004, Mod. Phys. Lett. A 19, 1745 ;  Sadeghi, J. et al., 2009, Phys. Rev. D 79, 123003.

\bibitem[Capozziello et al. (2003)]{cap03}  Capozziello, S., Carloni, S.  \&  Troisi, A., 2003, Recent Res. Dev. Astron. Astrophys. 1,  625.

\bibitem[Nojiri \&  Odintsov (2003)]{noj03} Nojiri, S. \&  Odintsov, S.D.2003 Phys. Rev. D 68, 123512; Phys. Lett. B 576, 5.

\bibitem[(Faraoni 2007, de Souza \&  Faraoni 2007, Brookfield et al. 2006, Briscese et al. 2007)]{far07}  Faraoni, V. 2007 Phys. Rev. D 75 067302; de Souza,J.C.C. \&  Faraoni,V. 2007, Class. Quant. Grav. 24 3637;  Brookfield,A.W. et al., 2006,  Phys. Rev. D 74 064028;  Briscese, F. et al., 2007, Phys. Lett. B 646 105.

\bibitem[(Farajollahi \&  Salehi 2010, Clifton  \&  Barrow 2005, Srivastava 2007, Nojiri et al. 2007, Baghram et al. 2007, Farajollahi \&  Milani 2010)]{far10} Farajollahi, H. \&  Salehi, A. 2010 J. Cosmol. Astropart. Phys. 11, 006; Clifton T. \&  Barrow,J. D. 2005, Phys. Rev. D 72 103005 ;  Koivisto,T. 2007, Phys. Rev. D 76 043527 (2007) ;
Srivastava, S.K. 2007, Phys. Lett. B 648 119 ;  Nojiri,S. et al., 2007, Phys. Lett. B 651 22 4; Baghram, S. et al., 2007 Phys. Rev. D 75 044024 ;  Farajollahi,H. \&  Milani,F. 2010, Mod. Phy. lett. A, Vol. 25, No. 27 2349-2362 .

\bibitem[( Nojiri \&  Odintsov 2004,Cognola et al. 2005,2006, Henttunen et al. 2008)]{noj04}  Nojiri, S. \&  Odintsov, S.D.2004, Gen. Rel. Grav. 36 1765 ; Phys. Lett. B 599 137 ;  Cognola,G. et al., 2005, J. Cosmol. Astropart. Phys. 0502 010; 2006, Phys. Rev. D 73 084007;  Henttunen, K., Multamaki, T., \&  Vilja, I., 2008, Phys. Rev. D 77 024040.

\bibitem[(Setare \&  Jamil 2010, Davis et al. 2009, Ito \&  Nojiri 2009, Tamaki \& Tsujikawa 2008,Farajollahi \& Salehi 2010 )]{set10} Setare,M. R. \&  Jamil,M. 2010, Phys. Lett. B 690  1-4 ;  Davis,A. C., Schelpe, C. A.O., Shaw, D. J., 2009, Phys. Rev. D 80 064016 ;
 Ito, Y. \&  Nojiri, S. 2009, Phys. Rev. D 79:103008; Tamaki,T. \& Tsujikawa,S. 2008, Phys. Rev. D 78 084028 ; Farajollahi, H. \& Salehi, A. 2010 Int. J. Mod. Phys. D19:621-633.




\bibitem[(Mota \& Shaw 2007)]{mot07} Mota,D.F. \&  Shaw, D.J. 2007, Phys. Rev. D 75, 063501.

\bibitem[(Dimopoulos \& Axenides 2005)]{dim05} Dimopoulos, K. \& Axenides, M. 2005, J. Cosmol. Astropart. Phys. 0506:008.


\bibitem[Jacobson (1995)]{jac95} Jacobson,T. 1995, Phys. Rev. Lett. 75 1260.

\bibitem[(Padmanabhan 2002, 2005)]{pad02} Padmanabhan,T. 2002, Class. Quant. Grav. 19 5387;2005, Phys. Rep. 406 49.

\bibitem[(Frolov \& Kofman 2003)]{fro03} Frolov, A. V. \& Kofman, L. 2003,  J. Cosmol. Astropart. Phys. 0305 009.

\bibitem[(Danielsson 2005)]{dan05}Danielsson, U. H. 2005, Phys. Rev. D 71 023516.

\bibitem[Bousso (2005)]{bou05} Bousso, R. 2005,  Phys. Rev. D 71 064024.

\bibitem[(Akbar \& Cai 2007)]{akb07} Akbar M. \& Cai, R. G. 2007, Phys. Rev. D 75 084003.

\bibitem[( Cai \& Cao 2007)]{cai07} Cai, R. G. \& Cao, L. M. 2007, Phys. Rev. D 75 064008; Nucl. Phys. B 785 135.

\bibitem[(Sheykhi et al. 2007, )]{she07} Sheykhi,A. et al. 2007, Nucl. Phys. B 779 1; Phys. Rev. D 76 023515;

\bibitem[Eling et al.(2006)]{eli06}Eling, C., Guedens, R. \& Jacobson, T., 2006, Phys. Rev. Lett. 96 121301.

\bibitem[Akbar \&  Cai (2006)]{akb06}Akbar, M. \&  Cai, R. G. 2006, Phys. Lett. B 635 7; 2007, Phys. Lett. B 648 243.

\bibitem[(Nojiri \& Odintsov 2004, Mohseni Sadjadi 2007)]{noj04} Nojiri, S. \& Odintsov, S. D., 2004, Phys. Rev. D 70, 103522;  Mohseni Sadjadi, H., 2007, J. Cosmol. Astropart. Phys. 0702 026.

\bibitem[Cai \&  Kim (2005)] {cai05} Cai, R. G. \&  Kim, S. P., 2005, JHEP 02 050.

\bibitem[( Wang 2005, Akbar \& Cai 2006)]{wan05} Wang,P., 2005, Phys. Rev. D 72, 024030.


\bibitem[(Setare \& Shafei 2006, Setare 2007, Davies 1987, Pollock \&  Singh 1989, Pavon 1990, Mukohyama 1997, Brustein 2000, Davis, T. M. et al. 2003, Izquierdo \&  Pavon 2006, Setare 2006, Mohseni Sadjadi 2007, Gong et al. 2007, Horvat 2007, Setare \& Vagenas 2008, Jamil et al. 2010 )]{dav87} M. R. Setare, S. Shafei 2006, J. Cosmol. Astropart. Phys. 0609  011, M. R. Setare, J. Cosmol. Astropart. Phys. 2007, 0701 023 Davies,P. C. W., 1987, Classical Quantum Gravity 4, L255; 1988, ibid. 5, 1349; Pollock, M. D. \&  Singh, M. D., 1989, Classical Quantum Gravity 6, 901; Pavon, D., 1990, Classical Quantum Gravity 7, 487;
Mukohyama, S., 1997, Phys. Rev. D 56, 2192; Brustein, R., 2000, Phys. Rev. Lett. 84, 2072;
Davis, T. M., Davies, P. C. W. \& Lineweaver, C. H., 2003, Classical Quantum Gravity 20, 2753;
Izquierdo,  G. \&  Pavon, D., 2006, Phys. Lett. B 639, 420; Setare, M. R., 2006, Phys.Lett. B641 130-133;
Mohseni Sadjadi, H., 2007, Phys. Lett. B 645, 108; Gong, Y., Wang, B. \& Wang, A., 2007, Phys. Rev. D 75, 123516;
Horvat, R., 2007, Phys. Lett. B 648, 374; M. R. Setare \& E. C. Vagenas  2008, PLB 666 111-115; M. Jamil, E. N. Saridakis, M. R. Setare 2010, Phys. Rev. D 81 023007; J. Cosmol. Astropart. Phys. 1011 032



\bibitem[(Farajollahi et al. 2011)]{farajo}H. Farajollahi, A. Salehi, F. Tayebi, 2011 Astrophys. Space Sci. 335:629–634


\bibitem[Sahni \& Starobinsky(2000)]{sah00} Sahni, V. \& Starobinsky, A.A.2000, Int. J. Mod. Phys. D 9, 373.

\bibitem[Zhou et al.(2007)]{zho07} Zhou, J., Wang, B., Gong, Y. \& Abdalla, E., 2007,  Phys. Lett. B 652, 86.



\end{thebibliography}
\end{document}